\documentclass[12pt]{article}
\usepackage{amsmath}
\usepackage{amssymb}
\usepackage{epsfig}
\usepackage{verbatim}
\usepackage{fancybox}
\usepackage{color}
\newcommand{\mathswitchr}[1]{\relax\ifmmode{\mathrm{#1}}\else$\mathrm{#1}$\fi}
\newcommand{\Ei}{\mathop{\mathrm{Ei}}\nolimits}
\newcommand{\Was}{W\c as}
\newcommand{\FYFS}{F_{\mathrm YFS}}
\newcommand{\rQCED}{\mathswitchr{QCED}}
\usepackage{cite}
\usepackage{epic}
\begin{document}
%=======================================================================
\begin{titlepage}
\begin{flushright}
{\bf BU-HEPP-09-03}\\
{\bf April 2009}\\
\end{flushright}
%\vspace{0.05cm}
 
\begin{center}
{\Large HERWIRI1.0: MC Realization of IR-Improved DGLAP-CS Parton Showers$^{\dagger}$}
\end{center}

\vspace{2mm}
\begin{center}
%%  {\bf   S. Jadach$^{a,b}$ and B.F.L. Ward$^{c,d}$}
{\bf   S. Joseph$^a$, S. Majhi$^b$, B.F.L. Ward$^a$ and S.A. Yost$^c$}\\
\vspace{2mm}
%{\em $^a$CERN, Theory Division, CH-1211 Geneva 23, Switzerland,}\\
%{\em $^b$Institute of Nuclear Physics,
%        ul. Kawiory 26a, Krak\'ow, Poland,}
%{\em $^c$Werner-Heisenberg-Institut, Max-Planck-Institut fuer Physik,
%Muenchen, Germany,}\\
%{\em $^a$Werner-Heisenberg-Institut, Max-Planck-Institut fuer Physik,
%Muenchen, Germany,}\\
{\em $^a$Department of Physics,\\
 Baylor University, Waco, Texas, USA}\\
%{\em $^b$SLAC, Stanford University, Stanford, California 94309, USA,}\\
{\em $^b$Theory Division, Saha Institute of Nuclear Physics,\\
  Kolkata, India }\\
{\em $^c$Department of Physics, The Citadel, Charleston, South Carolina, USA}\\
\end{center}

\vspace{5mm}
\begin{center}
{\bf   Abstract}
\end{center}
We present Monte Carlo data showing the comparison between the parton shower
generated by the standard Dokshitzer-Gribov-Lipatov-Altarelli-Parisi-Callan-Symanzik (DGLAP-CS) kernels and that generated with the new
IR-improved DGLAP-CS kernels recently developed by one of us. We do this
in the context of HERWIG6.5 by implementing the new kernels therein to generate a new MC, HERWIRI1.0, for hadron-hadron interactions at high energies.
We discuss possible phenomenological implications for precision
LHC theory. We also present comparisons with FNAL data.
\\
\vskip 20mm
\vspace{10mm}
\renewcommand{\baselinestretch}{0.1}
\footnoterule
\noindent
{\footnotesize
\begin{itemize}
\item[${\dagger}$]
Work partly supported by US DOE grants DE-FG02-05ER41399,  
DE-FG02-09ER41600 and 
% the Polish Government
%grants KBN 2P30225206 and 2P03B17210, the Maria Sk\l{}odowska-Curie
%Joint Fund II PAA/DOE-97-316, and
by NATO Grant PST.CLG.980342.
%, and by
%Polish Government grant 5P03B09320.
\end{itemize}
}
%\vspace{0.5cm}
%\begin{flushleft}
%{\bf UTHEP-00-0101}\\
%{\bf Jan, 2000}\\
%\end{flushleft}

\end{titlepage}

%=======================================================================
\def\Kmax{K_{\rm max}}\def\ieps{{i\epsilon}}\def\rQCD{{\rm QCD}}
\renewcommand{\theequation}{\arabic{equation}}
\font\fortssbx=cmssbx10 scaled \magstep2
\renewcommand\thepage{}
%\vfill\eject
\parskip.1truein\parindent=20pt\pagenumbering{arabic}

\section{Introduction}\label{intro}

With the advent of the LHC, we enter the era of precision QCD, which 
is characterized by
predictions for QCD processes at the total precision~\cite{jadach1} tag~\footnote{By total precision of a theoretical prediction, we mean the technical and physical precisions combined in quadrature or otherwise, as appropriate.} of $1\%$ or better.
At such a precision as we have as our goal, issues such as the role of QED~\cite{qedeffects,radcor-ew} are an integral part of the discussion 
and we deal with this by the simultaneous
resummation of QED and QCD large infrared(IR) 
effects, $QED\otimes QCD$ resummation
~\cite{qced} in the presence of parton showers, to be realized on an 
event-by-event basis by MC methods. We stress that, as shown in
Refs.~\cite{radcor-ew}, no precision prediction for a hard LHC process
at the 1\% level can be complete without taking the large EW corrections into account. 

In proceeding with our
discussion, we first review
our approach to resummation and its relationship to those in Refs.~\cite{cattrent,scet}; this review is 
followed by a summary of the attendant new IR-improved~\cite{irdglap1,irdglap2} DGLAP-CS theory~~\cite{dglap,cs} with some discussion of its implications.
We then present the implementation of the new IR-improved kernels in the framework of HERWIG6.5~\cite{herwig} to arrive at the new, IR-improved parton shower MC
HERWIRI1.0. We illustrate the effects of the IR-improvement first with the 
generic 2$\rightarrow$2 processes at LHC energies and then  with the specific
single $Z$ production process at LHC energies. The IR-improved showers are generally softer as expected and we discuss possible implications for precision LHC physics. We compare with recent data from FNAL to make
direct contact with observation. 
Section~\ref{concl} contains our summary remarks.

To put the discussion in the proper perspective, we note that the authors in
Ref.~\cite{scott1,scott2} have argued that
the current state-of-the-art theoretical precision tag on single $Z$
production at the LHC is 
$(4.1 \pm 0.3)\% = (1.51 \pm 0.75)\%$ (QCD) $ \oplus\ 3.79\%$ (PDF)
$ \oplus\ 0.38 \pm 0.26\% $ (EW) and that the analogous estimate for single $W$ 
production is $\sim 5.7$\%. These estimates, which can be considered as 
lower bounds, show how much work is still needed to achieve the desired 
1.0\% total precision tag on these two processes, for example. 
This point cannot be over-emphasized.
%,
%where the results of Refs.~\cite{cteq,mrst,mcnlo,fewz,resbos,horace,photos} have been
%used.
%\footnote{Recently, the 
%analogous estimate for single $W$ production has been given ~\cite{scott2} as $\sim 5.7$\%.}\par%\vskip0.2cm
%%%Start Here

\section{QED$\otimes$QCD Resummation}
In Refs.~\cite{qced,irdglap1,irdglap2}, we have derived the following expression for the 
hard cross sections in the SM $SU_{2L}\times U_1\times SU_3^c$ EW-QCD theory
\begin{eqnarray}
d\hat\sigma_{\rm exp} &=& e^{\rm SUM_{IR}(QCED)}
   \sum_{{n,m}=0}^\infty\frac{1}{n!m!}\int
\frac{d^3p_2}{p_2^{\,0}}\frac{d^3q_2}{q_2^{\,0}}
\prod_{j_1=1}^n\frac{d^3k_{j_1}}{k_{j_1}} 
\prod_{j_2=1}^m\frac{d^3{k'}_{j_2}}{{k'}_{j_2}}
\nonumber\\
& & \kern-2cm \times \int\frac{d^4y}{(2\pi)^4}
e^{iy\cdot(p_1+q_1-p_2-q_2-\sum k_{j_1}-\sum {k'}_{j_2})+ D_\rQCED}
\ \tilde{\bar\beta}_{n,m}(k_1,\ldots,k_n;k'_1,\ldots,k'_m),
\label{subp15b}
\end{eqnarray}
where the new YFS-style~\cite{yfs} residuals
$\tilde{\bar\beta}_{n,m}(k_1,\ldots,k_n;k'_1,\ldots,k'_m)$ have $n$ hard gluons and $m$ hard photons and we show the final state with two hard final
partons with momenta $p_2,\; q_2$ specified for a generic 2f final state for
definiteness. The infrared functions ${\rm SUM_{IR}(QCED)},\; D_\rQCED\; $
are defined in Refs.~\cite{qced,irdglap1,irdglap2}. This is the simultaneous resummation of QED and QCD large IR effects. Eq.\ (\ref{subp15b}) is an exact
implementation of amplitude-based resummation of the latter
effects valid to all orders in $\alpha$ and in $\alpha_s$.

Our approach to QCD resummation is fully consistent with that of
Refs.~\cite{cattrent,scet} as follows. First, Ref.~\cite{geor1} has shown that the latter two approaches are equivalent. We show in Refs.~\cite{irdglap1,irdglap2}
that our approach is consistent with that of Refs.~\cite{cattrent}
by exhibiting the transformation prescription from the resummation formula
for the theory in Refs.~\cite{cattrent} for the generic $2\rightarrow n$ parton process as given in Ref.~\cite{madg} to our theory as given for QCD by restricting Eq.(\ref{subp15b}) to its QCD component, where a key point is to use the color-spin density matrix formulation of our residuals to capture the respective full quantum mechanical color-spin correlations in the results in Ref.~\cite{madg} -- see Refs.~\cite{irdglap1,irdglap2} for details.

%\section{IR-Improved DGLAP-CS Theory}
We show in Refs.~\cite{irdglap1,irdglap2} that the result Eq.(\ref{subp15b})
allows us to improve in the IR regime \footnote{This 
should be distinguished from the also important
resummation in parton density evolution for the ``$z\rightarrow 0$'' regime,
where Regge asymptotics obtain -- see for example Ref.~\cite{ermlv,guido}. This
improvement must also be taken into account for precision LHC predictions.} 
the kernels in DGLAP-CS~\cite{dglap,cs}
theory as follows, using a standard notation:
\begin{align}
P^{exp}_{qq}(z)&= C_F \FYFS(\gamma_q)e^{\frac{1}{2}\delta_q}\left[\frac{1+z^2}{1-z}(1-z)^{\gamma_q} -f_q(\gamma_q)\delta(1-z)\right],\nonumber\\
P^{exp}_{Gq}(z)&= C_F \FYFS(\gamma_q)e^{\frac{1}{2}\delta_q}\frac{1+(1-z)^2}{z} z^{\gamma_q},\nonumber\\
P^{exp}_{GG}(z)&= 2C_G \FYFS(\gamma_G)e^{\frac{1}{2}\delta_G}\{ \frac{1-z}{z}z^{\gamma_G}+\frac{z}{1-z}(1-z)^{\gamma_G}\nonumber\\
&\qquad +\frac{1}{2}(z^{1+\gamma_G}(1-z)+z(1-z)^{1+\gamma_G}) - f_G(\gamma_G) \delta(1-z)\},\nonumber\\
P^{exp}_{qG}(z)&= \FYFS(\gamma_G)e^{\frac{1}{2}\delta_G}\frac{1}{2}\{ z^2(1-z)^{\gamma_G}+(1-z)^2z^{\gamma_G}\},
%P_{qG}(z)&=\frac{1}{2}(z^2+(1-z)^2).
\label{dglap19}
\end{align}
where the superscript ``exp'' indicates that the kernel has been resummed as
predicted by Eq.(\ref{subp15b}) when it is restricted to QCD alone and where
\begin{align}
\gamma_q &= C_F\frac{\alpha_s}{\pi}t=\frac{4C_F}{\beta_0}, \qquad \qquad
\delta_q =\frac{\gamma_q}{2}+\frac{\alpha_sC_F}{\pi}(\frac{\pi^2}{3}-\frac{1}{2}),\nonumber\\
f_q(\gamma_q)&=\frac{2}{\gamma_q}-\frac{2}{\gamma_q+1}+\frac{1}{\gamma_q+2},\nonumber\\
\gamma_G &= C_G\frac{\alpha_s}{\pi}t=\frac{4C_G}{\beta_0}, \qquad \qquad
\delta_G =\frac{\gamma_G}{2}+\frac{\alpha_sC_G}{\pi}(\frac{\pi^2}{3}-\frac{1}{2}),\nonumber\\
f_G(\gamma_G)&=\frac{n_f}{6C_G \FYFS(\gamma_G)}{e^{-\frac{1}{2}\delta_G}}+
\frac{2}{\gamma_G(1+\gamma_G)(2+\gamma_G)}+\frac{1}{(1+\gamma_G)(2+\gamma_G)},\\
%+\frac{1}{12}\}.
&\qquad +\frac{1}{2(3+\gamma_G)(4+\gamma_G)}+\frac{1}{(2+\gamma_G)(3+\gamma_G)(4+\gamma_G)},\nonumber\\
\FYFS(\gamma)&=\frac{e^{-C\gamma}}{\Gamma(1+\gamma)}, \qquad \qquad \qquad C=0.57721566... ,
\end{align}
where $\Gamma(w)$ is Euler's gamma function and $C$ is Euler's constant.
We use a one-loop formula for $\alpha_s(Q)$, so that
\[\beta_0=11-\frac{2}{3}n_f,\] where $n_f$ is the number of
active quark flavors and $C_F=4/3$ and $C_G=3$ are the 
respective quadratic Casimir invariants 
for the quark and gluon color representations
-- see Refs.~\cite{irdglap1,irdglap2} for the corresponding details. 
The results in Eq.(\ref{dglap19}) have now been implemented by 
MC methods, as we exhibit in the following sections.

\section{Illustrative Results/Implications}
Firstly, we note that the connection to the higher order kernels in Refs.~\cite{high-ord-krnls} has been made in Ref.\ \cite{irdglap1}. This opens 
the way for the systematic improvement of the results presented herein.
Secondly, in the NS case, we find~\cite{irdglap1} that the $n=2$ moment
is modified by $\sim 5\%$ when evolved with Eq.(\ref{dglap19}) 
from $2$GeV to $100$GeV with $n_f=5$
and $\Lambda_{QCD}\cong 0.2GeV$, for illustration. This effect is thus relevant
to the expected precision of the HERA final data analysis~\cite{hera-dat}.
Thirdly, we have been able to use
Eq.(\ref{subp15b}) to resolve the violation~\cite{sac-no-go,cat1} 
of Bloch-Nordsieck cancellation in 
ISR(initial state radiation) 
at ${\cal O}(\alpha_s^2)$ for massive quarks~\cite{qmass-bw}.
This opens the way to include realistic quark masses as we introduce the
higher order EW corrections in the presence of higher order QCD corrections 
-- note that the radiation probability in QED at the hard scale $Q$ involves 
the logarithm $\ln(Q^2/m_q^2)$, and it will not do to set $m_q=0$ to analyze 
these effects in a fully exclusive, differential event-by-event calculation 
of the type that we are constructing. 
Fourthly, the threshold resummation implied by Eq.(\ref{subp15b}) for single $Z$
production at LHC shows a $0.3\%$ QED effect and agrees with known exact
results in QCD -- see Refs.~\cite{qced,baurall,exactqcd}. Fifthly, we have a 
new scheme~\cite{irdglap2} for precision LHC theory: in an obvious notation,
\begin{equation}
%\begin{split}
\sigma =\sum_{i,j}\int dx_1dx_2F_i(x_1)F_j(x_2)\hat\sigma(x_1x_2s)
       =\sum_{i,j}\int dx_1dx_2{F'}_i(x_1){F'}_j(x_2)\hat\sigma'(x_1x_2s),
%\end{split}
\label{sigscheme}
\end{equation}
where the primed quantities are associated with Eq.(\ref{dglap19}) in the
standard QCD factorization calculus. Sixthly, we have~\cite{qced} an attendant
shower/ME matching scheme, wherein, for example, in combining Eq.(\ref{subp15b})
with HERWIG~\cite{herwig}, PYTHIA~\cite{pythia}, MC@NLO~\cite{mcnlo}
or new shower MC's~\cite{skrzjad}, we may use either
$p_T$-matching
or shower-subtracted residuals\newline $\{\hat{\tilde{\bar\beta}}_{n,m}(k_1,\ldots,k_n;k'_1,\ldots,k'_m)\}$ to create a paradigm without double
counting that can be systematically improved order-by order in
perturbation theory -- see Refs.~\cite{qced}. 

The stage is set for the full MC implementation of our approach. We turn next to the initial stage of this implementation -- that of the kernels in Eq.(\ref{dglap19}).

\section{MC Realization of IR-Improved DGLAP-CS\\
            Theory}
In this section we describe the initial implementation of the 
new IR-improved kernels in the HERWIG6.5 environment, which then results
in a new MC, which we denote by HERWIRI1.0, which stands for ``high energy radiation with IR improvement.''\footnote{We thank M. Seymour and B. Webber for discussion on this point.}

Specifically, our approach can be summarized as follows.
We modify the kernels in the HERWIG6.5 module HWBRAN and in the attendant
 related modules~\cite{bw-ms-priv} with the following substitutions:
\begin{equation}\text{DGLAP-CS}\; P_{AB}  \Rightarrow \text{IR-I DGLAP-CS}\; P^{exp}_{AB}
\label{substitn}
\end{equation}
while leaving the hard processes alone for the moment. We have in 
progress~\cite{inprog} %% (SY,BFLW,MH,SM,SJ)
the inclusion of YFS synthesized electroweak  
modules from Refs.~\cite{jad-ward} %{\Color{Magenta}Jadach et al.  MC's} 
for
HERWIG6.5, HERWIG++~\cite{herpp} hard processes. The fundamental issue is that
CTEQ~\cite{cteq} and MRST (MSTW after 2007)~\cite{mrst} best parton densities
do not include precision electroweak higher order corrections and such effects do enter in a 1\% precison tag budget for processes such as single heavy gauge boson production in the LHC environment, as we have emphasized. 

%%\end{itemize}
\def\beqa{\begin{eqnarray}}
\def\eeqa{\end{eqnarray}}
\def\beq{\begin{equation}}
\def\eeq{\end{equation}}
\def\non{\nonumber}
\def\no{\noindent }

For definiteness, let us illustrate the implementation by an example~\cite{bw-ann-rev,sjosback}, which for pedagogical reasons we will take as a simple leading
log shower component with a virtuality evolution variable, with the understanding that in HERWIG6.5 the shower development is angle ordered~\cite{bw-ann-rev} so that the evolution variable is actually $\sim E\theta$ where $\theta$ is the opening angle of the shower as defined in Ref.~\cite{bw-ann-rev} for a parton initial energy $E$. In this pedagogical example, which we take from Ref.~\cite{bw-ann-rev}, 
%\titbox{\Color{Maroon} Implementation Illustration}
the probability that no branching occurs above virtuality
cutoff $Q_0^2$ is  $\Delta_a(Q^2,Q_0^2)$ so that
%${\Color{Red}\Rightarrow}$
\beq \label{eq:splitprob}
d\Delta_a(t,Q_0^2) = \frac{-dt}{t}\Delta(t,Q_o^2)\sum_b\int dz\frac{\alpha_s}{2 \pi}P_{ba}(z),
\eeq
%\no
%${\Color{Red}\Rightarrow}$
which implies
\beq
\Delta_a(Q^2,Q_0^2)=\exp\left[ -\int_{Q_0^2}^{Q^2} \frac{dt}{t} \sum_b\int dz\frac{\alpha_s}{2 \pi}P_{ba}(z)\right].
\label{delta-a} 
\eeq
The attendant non-branching probability appearing in the evolution equation is
\beq
\Delta(Q^2,t) = \frac{\Delta_a(Q^2,Q_o^2)}{\Delta_a(t,Q_o^2)}, \quad t =k_a^2 \quad \text{the virtuality of gluon $a$}.
\eeq
The respective virtuality of parton $a$ is then generated with
\beq
\Delta_a(Q^2,t) = R,
\eeq
where $R$ is a random number uniformly distributed in $[0,1]$ .
With (note $\beta_0=b_0|_{n_c=3}$ here, where $n_c$ is the number of colors)
\beqa
\alpha_s(Q) = \frac{2 \pi}{b_0 \log\left(\frac{Q}{\Lambda}\right)},
\eeqa
we get for example 
\beqa
\int_0^1 dz \frac{\alpha_s(Q^2)}{2 \pi} P_{qG}(z)
&=& \frac{4\pi}{2 \pi b_0\ln\left(\frac{Q^2}{\Lambda^2}\right)}\int_0^1 dz \frac{1}{2}\left[ z^2+(1-z)^2\right] \non\\ 
&=& \frac{2}{3} \frac{1}{b_0\ln\left(\frac{Q^2}{\Lambda^2}\right)}.
\eeqa
so that the subsequent integration over $dt$ yields 
%${\Color{Red}\Rightarrow}$
\beqa
&&I=\int_{Q_0^2}^{Q^2}\frac{1}{3} \frac{dt}{t}\frac{2}{ b_0 \ln\left(\frac{t}{\Lambda^2}\right)} \non \\
%,\quad t=Q^2 \non \\
&=& \frac{2}{3b_0}\ln \ln \frac{t}{\Lambda^2}|_{Q_0^2}^{Q^2} \non \\
&=& \frac{2}{3 b_0}\left[\ln \left(\frac{\ln\left(\frac{Q^2}{\Lambda^2}\right)}{\ln\left(\frac{Q_0^2}{\Lambda^2}\right)}\right)\right].
\eeqa

Finally, introducing $I$ into Eq.(\ref{delta-a}) yields 
\beqa \label{DeltaQHerwig}
\Delta_a(Q^2,Q_0^2) &=& \exp \left[-\frac{2}{3 b_0}\ln \left(\frac{\ln\left(\frac{Q^2}{\Lambda^2}\right)}{\ln\left(\frac{Q_0^2}{\Lambda^2}\right)}\right)\right]\non\\
&=& \left[\frac{\ln\left(\frac{Q^2}{\Lambda^2}\right)}{\ln\left(\frac{Q_0^2}{\Lambda^2}\right)}\right]^{-\frac{2}{3b_0}}.
\eeqa
If we now let
$\Delta_a(Q^2,t)=R$, then
\beq 
\left[\frac{\ln\left(\frac{t}{\Lambda^2}\right)}{\ln\left(\frac{Q^2}{\Lambda^2}\right)}\right]^{\frac{2}{3b_0}} = R
\eeq
which implies
\beq
t = \Lambda^2 \left(\frac{Q^2}{\Lambda^2}\right)^{R^{\frac{3 b_0}{2}}}.
\label{t-herwig}
\eeq
Recall in HERWIG6.5~\cite{herwig} we have 
\beqa
b_0 &=& \left(\frac{11}{3}n_c - \frac{2}{3}n_f\right) \non \\ 
&=& \frac{1}{3}\left(11n_c - 10 \right), \quad n_f =5 \non \\
&\equiv& \frac{2}{3} {\tt BETAF}
\eeqa
where in the last line we used the notation in HERWIG6.5. 
%Note the
%simple relation $\beta_0=b_0|_{n_c=3}$ in our notations.
The momentum available after a  $q\bar{q}$ split in HERWIG6.5~\cite{herwig} 
is given by
\beq
{\tt QQBAR} = {\tt QCDL3} \left(\frac{\tt QLST}{\tt QCDL3}\right)^{R^{\tt BETAF}},
\eeq
in complete agreement with Eq.(\ref{t-herwig}) when we note the
identifications $t={\tt QQBAR}^2,\;\Lambda\equiv {\tt QCDL3},
\; Q\equiv {\tt QLST}$.

The leading log exercise leads to the same algebraic relationship that
HERWIG6.5 has between {\tt QQBAR} and {\tt QLST} but 
we stress that in HERWIG6.5
these quantities are the angle-ordered counterparts of the virtualities 
we used in our example, so that the shower is angle-ordered.

\begin{figure}[h]
\begin{center}
\scalebox{0.5}{\includegraphics[angle=90]{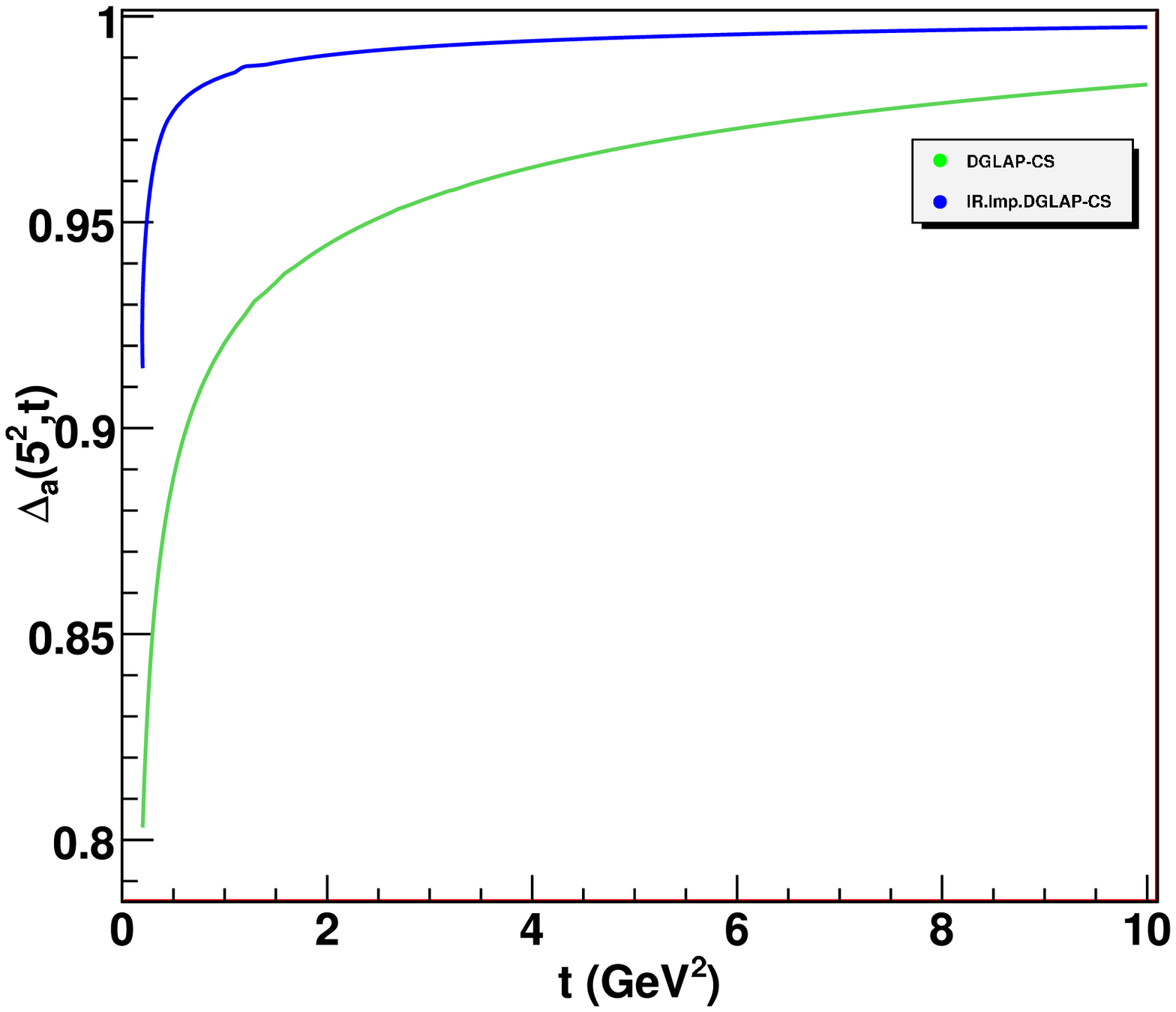}}
\end{center}
\caption{ Graph of $\Delta_a(Q^2,t)$ for the DGLAP-CS and IR-Improved DGLAP-CS kernels Eqs. (\ref{DeltaQHerwig}, \ref{DeltaQWard}). Q$^2$ is a typical virtuality close to the squared scale of the hard subprocess 
-- here we use $Q^2=25$GeV$^2$ for illustration.} 
\label{iri-vs-cs}
\end{figure}

Let us now repeat the above calculation for the IR-Improved kernels in 
Eq.(\ref{dglap19}). We have
\beq
P_{qG}^{\exp}(z) = \FYFS(\gamma_G)e^{\delta_G/2}\frac{1}{2} 
                    \biggl[ z^2(1-z)^{\gamma_G} + (1-z)^2z^{\gamma_G} \biggr]
\eeq
so that
\beq
\int_0^1 dz \frac{\alpha_s\left(Q^2\right)}{2 \pi} P_{qG}(z)^{\exp}
 = \frac{4\FYFS(\gamma_G)e^{\delta_G/2} }{b_0 \ln\left(\frac{Q^2}{\Lambda^2}\right)\left(\gamma_G+1\right)\left(\gamma_G+2\right)\left(\gamma_G+3\right)}.
\eeq
This leads to the following integral over $dt$
\beqa
&&I=\int_{Q_0^2}^{Q^2}\frac{dt}{t} \frac{4\FYFS(\gamma_G)e^{\delta_G/2} }{b_0 \ln\left(\frac{t}{\Lambda^2}\right)\left(\gamma_G+1\right)\left(\gamma_G+2\right)\left(\gamma_G+3\right)} \non\\
%,\quad t=Q^2 \non \\ 
&=&\frac{4 \FYFS(\gamma_G)e^{\gamma_G/4}}{b_0\left(\gamma_G+1\right)\left(\gamma_G+2\right)\left(\gamma_G+3\right)}
\Ei\left(1,\frac{8.369604402}{b_0\ln\left(\frac{t}{\Lambda^2}\right)}\right) \Bigg\vert_{Q_0^2}^{Q^2}.
\eeqa
Here we have used
\beq
\delta_G=\frac{\gamma_G}{2} + \frac{\alpha_s C_G}{\pi}\left(\frac{\pi^2}{3} - \frac{1}{2}\right),
\eeq
with $C_G=3$ the gluon quadratic Casimir invariant.
We finally get the IR-improved formula
\beq \label{DeltaQWard}
\Delta_a(Q^2,t) = \exp\left[-\left(F\left(Q^2\right)-F\left(t\right)\right)\right],
\eeq
where
\beq
F(Q^2) = \frac{4 \FYFS(\gamma_G)e^{\gamma_G/4}}{b_0\left(\gamma_G+1\right)\left(\gamma_G+2\right)\left(\gamma_G+3\right)}
\Ei\left(1,\frac{8.369604402}{b_0\ln\left(\frac{Q^2}{\Lambda^2}\right)}\right),
\eeq
and $\Ei$  is the exponential integral function.
In Fig.~\ref{iri-vs-cs} we show the difference between the two results for 
$\Delta_a(Q^2,t)$. We see that they agree within a few percent except for the softer values of $t$, as expected. We look forward to determining definitively
whether the experimental data prefer one over the other. This detailed study will appear elsewhere~\cite{elswh} but we begin the discussion below with a view on recent FNAL data. 
Again, we note that the comparison in Fig.~\ref{iri-vs-cs} is carried out at the leading log virtuality level, but the subleading effects suppressed in this 
discussion will not change our general conclusions drawn therefrom.

For further illustration, we note that
for the $q\rightarrow qG$ branching process in HERWIG6.5~\cite{herwig}, we have therein the implementation of the usual DGLAP-CS kernel as follows:

\begin{minipage}[c]{0.75\linewidth}
\begin{verbatim}
      WMIN = MIN(ZMIN*(1. -ZMIN), ZMAX*(1.-ZMAX))
      ETEST = (1. + ZMAX**2) * HWUALF(5-SUDORD*2, QNOW*WMIN)
      ZRAT = ZMAX/ZMIN
30    Z1 = ZMIN * ZRAT**HWRGEN(0)
      Z2 = 1. - Z1
      PGQW = (1. + Z2*Z2)
      ZTEST = PGQW * HWUALF(5-SUDORD*2, QNOW*Z1*Z2)
      IF (ZTEST .LT. ETEST*HWRGEN(1)) GOTO 30
      ...

\end{verbatim} 
\end{minipage}\begin{minipage}[c]{0.25\linewidth}
\begin{equation}
\hfill
\label{exhwb1}
\end{equation}
\end{minipage}
%\begin{align}
%\;\;\;\;\;&WMIN=MIN(ZMIN*(1. -ZMIN),ZMAX*(1.-ZMAX))\nonumber\\
%\;\;\;\;\;&ETEST=(1.+ZMAX**2)*HWUALF(5-SUDORD*2,QNOW*WMIN)\nonumber\\
%\;\;\;\;\;&ZRAT=ZMAX/ZMIN\nonumber\\
%\;\;\;30\;&Z1=ZMIN*ZRAT**HWRGEN(0)\nonumber\\
%\;\;\;\;\;&Z2=1.-Z1\nonumber\\
%\;\;\;\;\;&PGQW = (1.+Z2*Z2)\nonumber\\
%\;\;\;\;\;&ZTEST=PGQW*HWUALF(5-SUDORD*2,QNOW*Z1*Z2)\nonumber\\
%\;\;\;\;\;&IF(ZTEST.LT.ETEST*HWRGEN(1)) GOTO\; 30\nonumber\\
%\;\;\;\;\;&\ldots
%\label{exhwb1}
%\end{align}
where the branching of $q$ to $G$ at $z=${\tt Z1} occurs in the interval from
{\tt ZMIN} to {\tt ZMAX} set by the inputs to the program and the current 
value of the virtuality {\tt QNOW}, {\tt HWUALF} is the respective function 
for $\alpha_s$ in the program and {\tt HWRGEN(J)} are uniformly 
distributed random numbers on the interval from 0 to 1. It is seen that
Eq.(\ref{exhwb1}) is a standard MC realization of the unexponentiated DGLAP-CS kernel via
\begin{equation}
\alpha_s(Qz(1-z))P_{Gq}(z)=\alpha_s(Qz(1-z))\frac{1+(1-z)^2}{z}
\end{equation}
where the normalization is set by the usual conservation of probability.
To realize this with the IR-improved kernel, we make the replacement
of the code in Eq.(\ref{exhwb1}) with the lines 
\begin{minipage}[c]{0.75\linewidth}
\begin{verbatim}
      NUMFLAV = 5
      B0 = 11. - 2./3.*NUMFLAV
      L = 16./(3.*B0)
      DELTAQ = L/2 + HWUALF(5-SUDORD*2, QNOW*WMIN)*1.184056810
      ETEST = (1. + ZMAX**2) * HWUALF(5-SUDORD*2, QNOW*WMIN)
            * EXP(0.5*DELTAQ) * FYFSQ(NUMFLAV-1) * ZMAX**L
      ZRAT = ZMAX/ZMIN
30    Z1 = ZMIN * ZRAT**HWRGEN(0)
      Z2 = 1. - Z1
      DELTAQ = L/2 + HWUALF(5-SUDORD*2, QNOW*Z1*Z2)*1.184056810
      PGQW = (1. + Z2*Z2) * EXP(0.5*DELTAQ) * FYFSQ(NUMFLAV-1)
           * Z1**L
      ZTEST = PGQW * HWUALF(5-SUDORD*2, QNOW*Z1*Z2)
      IF (ZTEST .LT. ETEST*HWRGEN(1)) GOTO 30
      ...

\end{verbatim} 
\end{minipage}\begin{minipage}[c]{0.25\linewidth}
\begin{equation}
\hfill
\label{exhwb2}
\end{equation}
\end{minipage}
%\begin{align}
%\;\;\;\;\;\;&NUMFLAV=5\nonumber\\
%\;\;\;\;\;\;&b0 = 11. -2./3*NUMFLAV\nonumber\\
%\;\;\;\;\;\;&l = 16./(3*b0)\nonumber\\
%\;\;\;\;\;\;&DELTAq =l/2+HWUALF(5-SUDORD*2,QNOW*WMIN)*1.184056810\nonumber\\
%\;\;\;\;\;\;&ETEST=(1.+ZMAX**2)*HWUALF(5-SUDORD*2,QNOW*WMIN)\nonumber\\
%\;\;\;\;\&\;& \qquad     *exp(.5*DELTAq)*FYFSq(NUMFLAV-1)*ZMAX**l\nonumber\\
%\;\;\;\;\;\;&ZRAT=ZMAX/ZMIN\nonumber\\
%\;\;30\;\;\;&Z1=ZMIN*ZRAT**HWRGEN(0)\nonumber\\
%\;\;\;\;\;\;&Z2=1.-Z1\nonumber\\
%\;\;\;\;\;\;&DELTAq =l/2+HWUALF(5-SUDORD*2,QNOW*Z1*Z2)*1.184056810\nonumber\\
%\;\;\;\;\;\;&PGQW = (1. + Z2*Z2)*exp(.5*DELTAq)*FYFSq(NUMFLAV-1)\nonumber\\
%\;\;\;\;\&\;& \qquad     *Z1**l\nonumber\\
%\;\;\;\;\;\;&ZTEST=PGQW*HWUALF(5-SUDORD*2,QNOW*Z1*Z2)\nonumber\\
%\;\;\;\;\;\;&IF (ZTEST.LT.ETEST*HWRGEN(1)) GOTO\; 30\nonumber\\
%\;\;\;\;\;\;&\ldots
%\label{exhwb2}
%\end{align}
so that with the identifications $\gamma_q\equiv {\tt L},\; \delta_q\equiv 
{\tt DELTAQ},\; \FYFS(\gamma_q)\equiv {\tt FYFSQ(NUMFLAV-1)}$, we see 
that Eq.(\ref{exhwb2}) realizes the IR-improved DGLAP-CS kernel 
$P^{\exp}_{Gq}(z)$ via $\alpha_s(Qz(1-z))P^{\exp}_{Gq}(z)$ with the 
normalization again set by probability conservation. Continuing in this way, 
we have  carried out the corresponding changes for all of the kernels 
in Eq.(\ref{dglap19}) in the HERWIG6.5 environment, with its angle-ordered 
showers, resulting in the new MC, HERWIRI1.0(31),
in which the ISR parton showers have IR-improvement as given by
the kernels in Eq.\ (\ref{substitn}).~\footnote{\label{fna} In the original 
release of the program, we stated that the time-like parton showers had 
been completely IR-improved in a way that suggested the space-like parton 
showers had not yet been IR-improved at all. We subsequently
introduced release 1.02 in which the part of the 
space-like parton showers associated with 
HERWIG6.5's space-like module HWSGQQ for the space-like branching process
$G\rightarrow q\bar{q}$ was IR-improved. Recently, the remaining un-IR-improved
aspect of the space-like branching process, that in HERWIG6.5's space-like
module HWSFBR, 
as also been IR-improved in release 1.031. All of the results
in this paper were obtained using the latter release.}
%The module HWSGQQ is now IR-improved as well in the release HERWIRI1.02.
%The effect is small, as these considerations suggest: we see effects at a 
%level comparable to the errors on the MC data 
%in our plots when going from version 1.0 to 1.02.}
%The extension of our IR improvement to the space-like parton showers will appear elsewehere~\cite{elswh}. 
%%STARTHERE
We now illustrate some of the results we have obtained in comparing 
ISR showers in HERWIG6.5 and with those in HERWIRI1.031 (see footnote \ref{fna}) at LHC
%%to be added when FNAL figs ready: 
and at FNAL
energies, where some comparison with real data is also featured at the FNAL
energy.
%\titbox{\Color{Maroon}RESULTS}
Specifically, we compare the $z$-distributions, $p_T$-distributions, {\it etc.},
that result from the IR-improved and usual
DGLAP-CS showers in what follows.\footnote{Similar comparisons for PYTHIA and MC@NLO are in progress and we show some results with MC@NLO below.}

\begin{figure}
\begin{center}
\epsfig{file=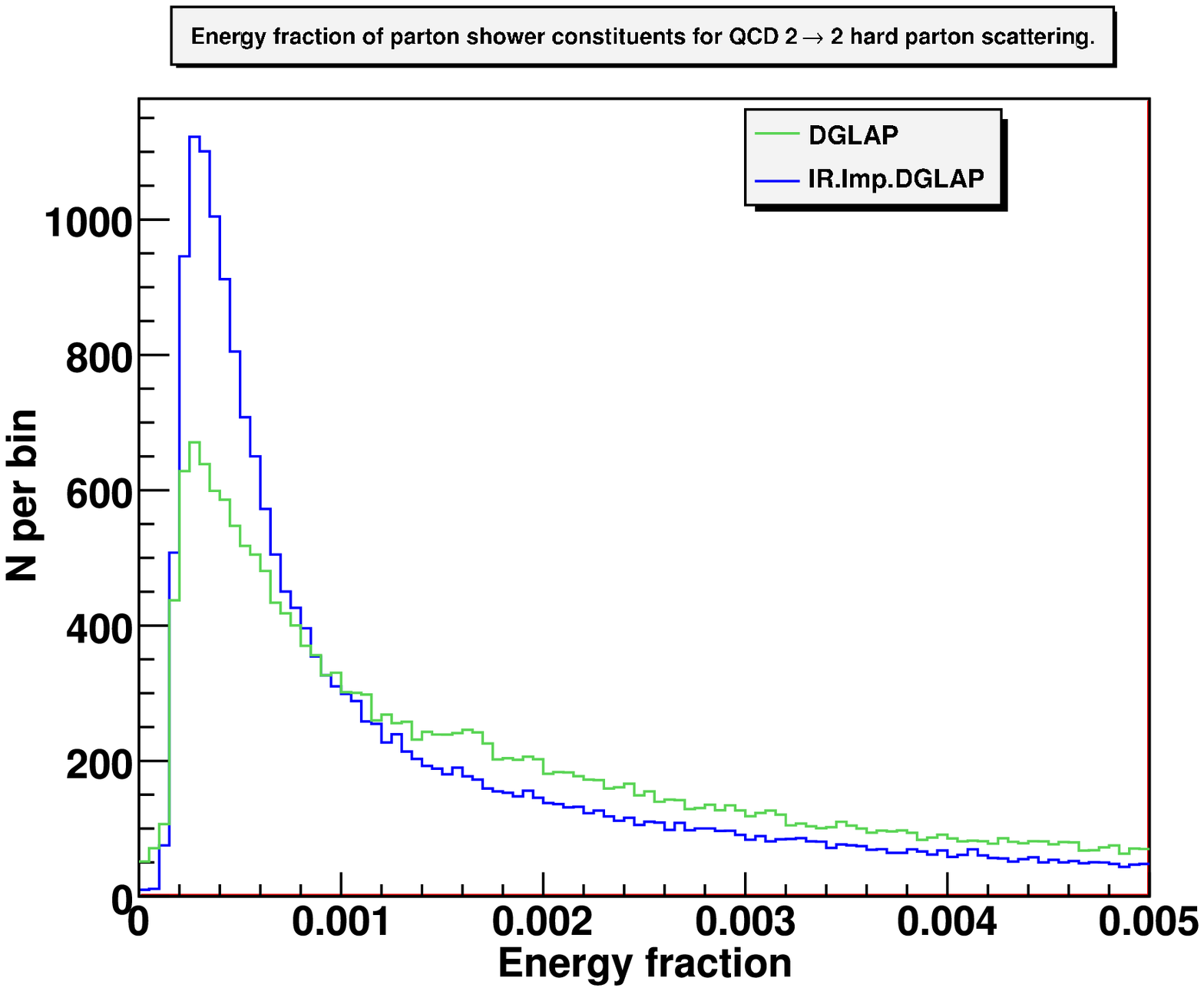,width=100mm,angle=90}
\end{center}
\caption{The $z$-distribution (ISR parton energy fraction) shower comparison in HERWIG6.5.}
\label{fighw1}
\end{figure}

\begin{figure}
\begin{center}
\epsfig{file=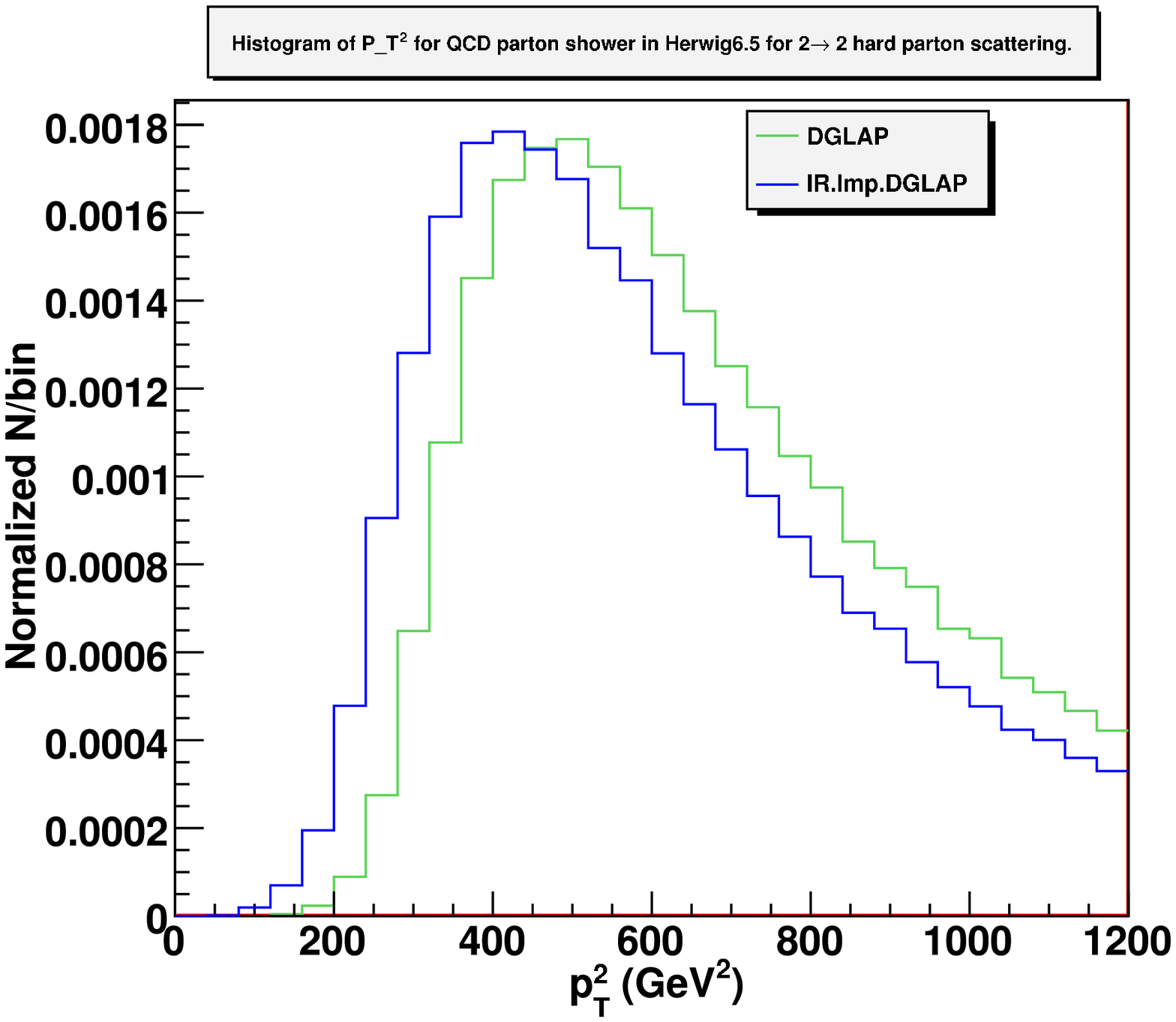,width=110mm,angle=90}
\end{center}
\caption{The $p_T^2$-distribution (ISR parton) shower comparison in HERWIG6.5.}
\label{fighw2}
\end{figure}

First, for the generic 2$\rightarrow$2 hard processes at LHC energies (14 TeV) we get the comparison shown Figs.~\ref{fighw1}, \ref{fighw2} for the respective ISR $z$-distribution and $p_T^2$ distribution at the parton level. 
Here, there are no cuts placed on the MC data and we define $z$ as
$z=E_{\text{parton}}/E_{\text{beam}}$ where $E_{\text{beam}}$ is the cms beam energy and $E_{\text{parton}}$ is the respective parton energy in the cms system. The two quantities $z$ and  $p_T^2$ for partons are of course not directly observable but their distributions show the softening of the IR divergence as we expect. 
  
%%\ref{fighw1},\ref{fighw2},\ref{fighw3}:
%%\ref{fighw1},\ref{fighw2},\ref{fighw3}:
\begin{figure}
\begin{center}
\epsfig{file=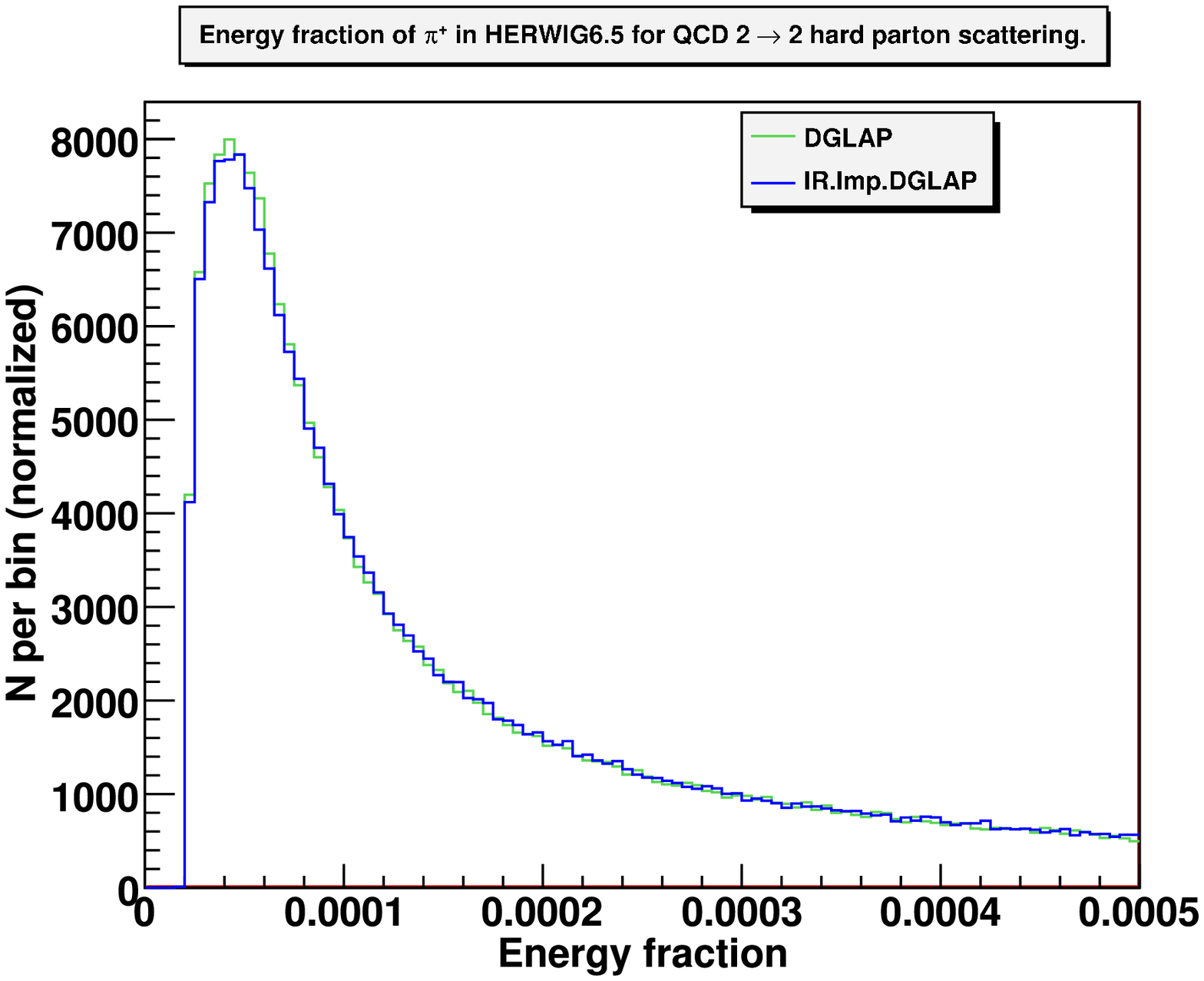,height=120mm,angle=90}
\end{center}
\caption{The $\pi^+$ energy fraction distribution shower comparison in HERWIG6.5.}
\label{fighw3}
\end{figure}

\begin{figure}
\begin{center}
\scalebox{0.7}{\includegraphics[angle=90]{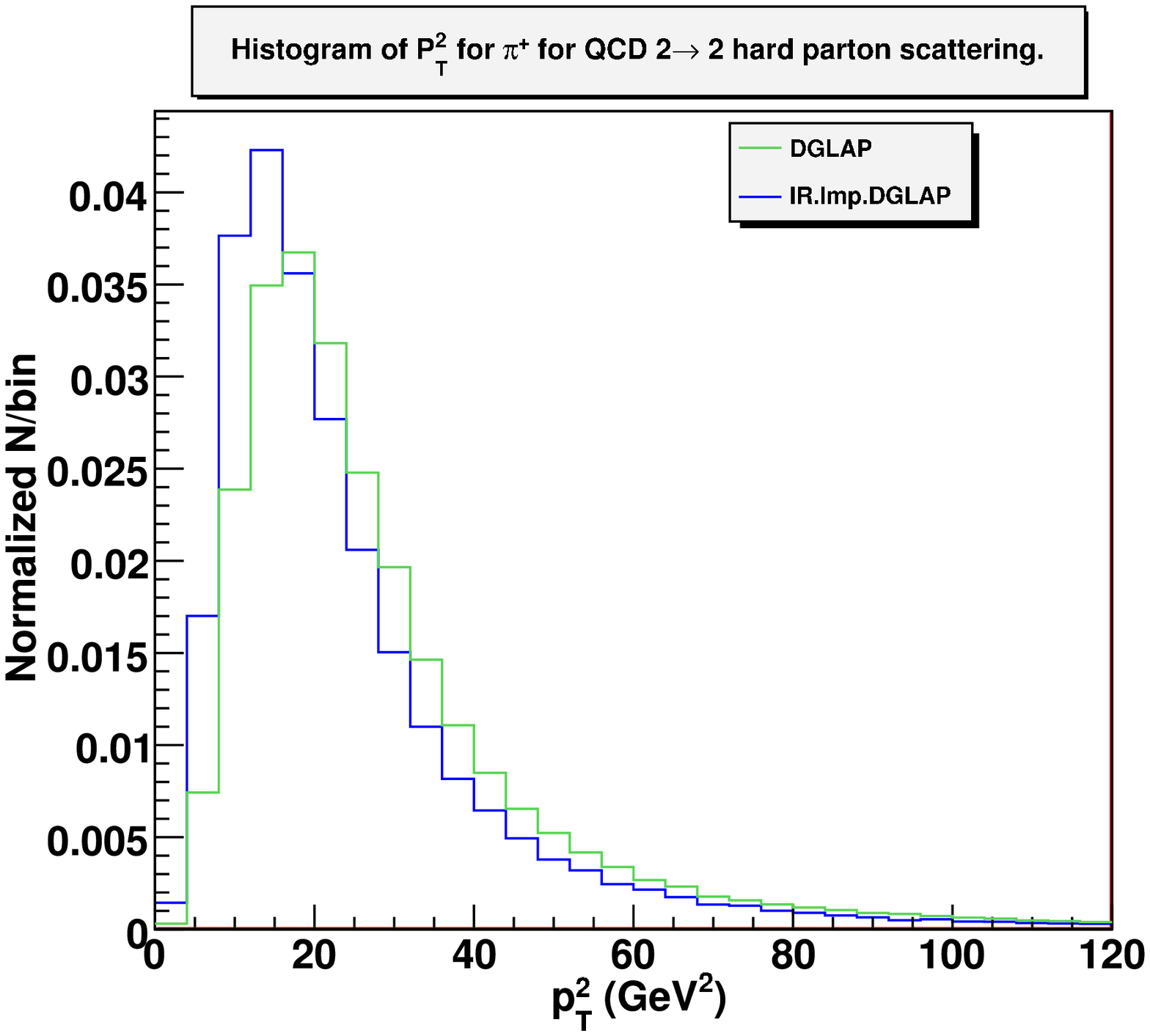}}
\end{center}
\caption{The $\pi^+$ $p_T^2$-distribution shower comparison in HERWIG6.5.}
\label{fighw4}
\end{figure}

%\ref{fighw1},\ref{fighw2},\ref{fighw3}:
Turning next to the similar quantities for the $\pi^+$ production in the 
generic 2$\rightarrow$2 hard processes at LHC, we see again in Figs.~\ref{fighw3}, \ref{fighw4} that the former spectra are very similar in the soft regime
while the latter spectra are softer in the IR-improved case. These spectra of course would be 
subject to some ``tuning'' in a real experiment and we await with anticipation the outcome of such 
an effort in comparison to LHC data.

\begin{figure}
\begin{center}
\epsfig{file=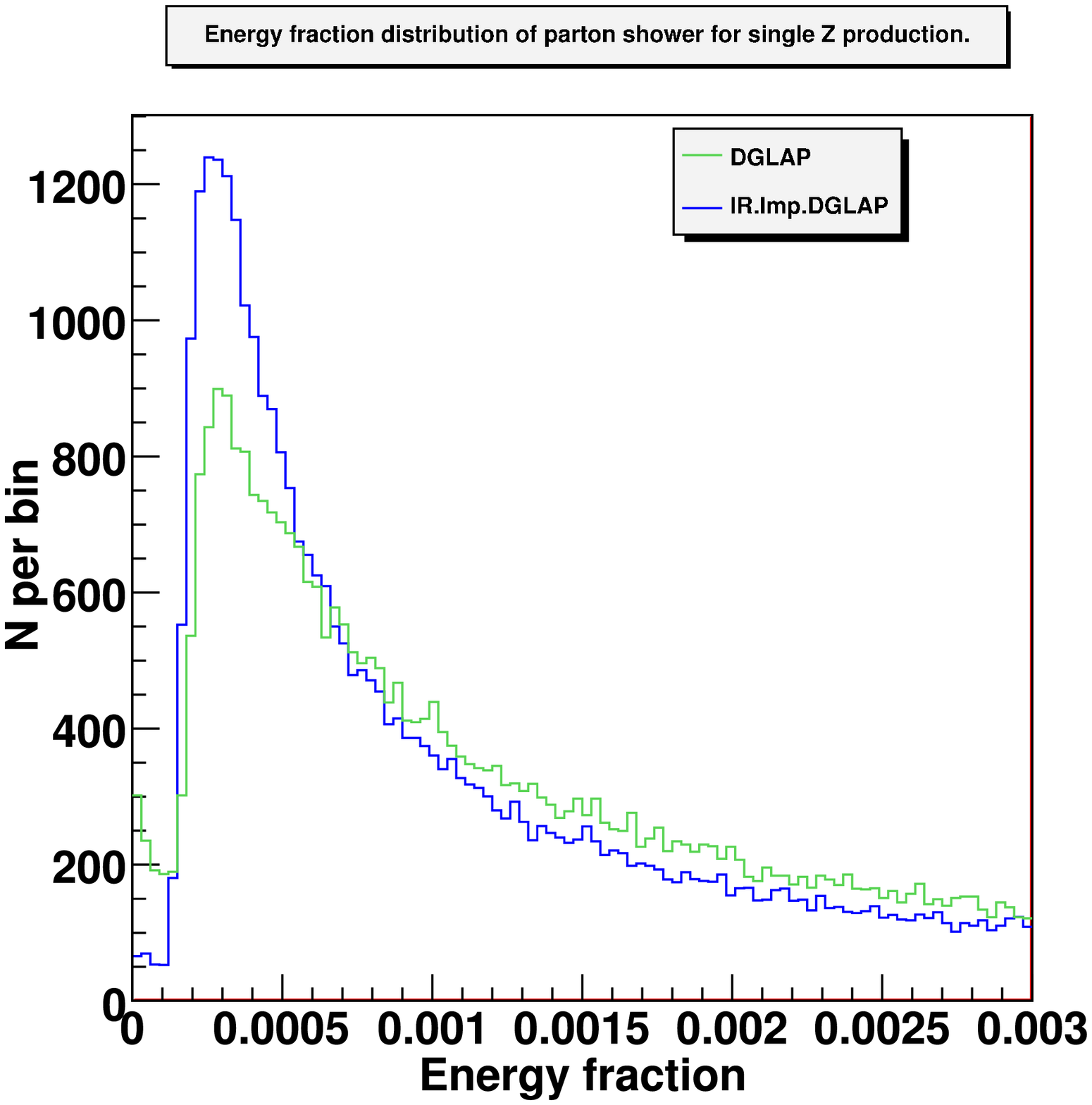,height=110mm,angle=90}
\end{center}
\caption{The $z$-distribution(ISR parton energy fraction) shower comparison in HERWIG6.5.}
\label{fighw5}
\end{figure}

\begin{figure}
\begin{center}
\epsfig{file=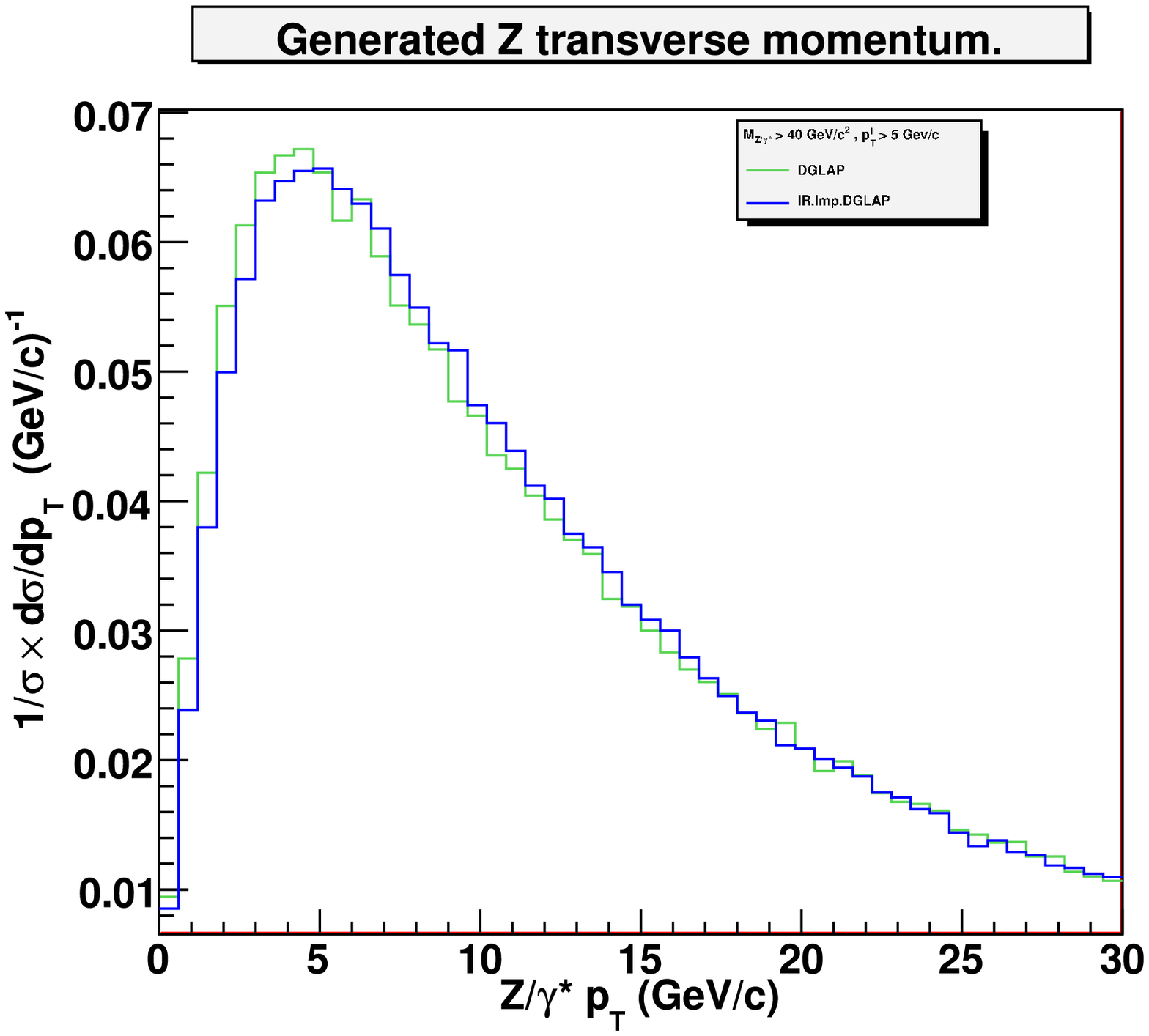,height=110mm,angle=90}
\end{center}
\caption{The $Z$ $p_T$-distribution(ISR parton shower effect) comparison in HERWIG6.5.}
\label{fighw6}
\end{figure}

We turn next to the luminosity process of single $Z$ production at the LHC, where in Figs.~\ref{fighw5} and \ref{fighw6} %,\ref{fighw7} 
we show respectively the ISR parton energy fraction distribution and 
the $Z$ $p_T$ distribution%, and the Z-rapidity distribution
with cuts on the acceptance as $40\text{GeV}<M_Z,\; p^\ell_T>5\text{GeV}$
% \; |\eta_\ell|<50$ 
for $Z\rightarrow \mu\bar{\mu}$ -- all lepton rapidities are included. For the energy fraction distribution we again see softer spectra in the IR-improved case
whereas for the $p_T$ distributions we see very similar spectra. 
%For the rapidity plot, we see the migration of some events to the higher values of $|\eta|$, again consistent with a softer spectrum for the IR-improved case.
We look forward to the confrontation with experiment,
where again we stress that in a real experiment, a certain amount of ``tuning'' will affect these results. We note for example that the difference between the
spectra in Fig.~\ref{fighw6}, while it is interesting, is well within
the range that could be tuned away by varying the amount of
intrinsic transverse momentum of partons in the proton. The question will always be which set of distributions gives a better $\chi^2$ per degree of freedom.

%\item Single Z-production at LHC
\begin{figure}
%\begin{center}
%%%%\epsfig{file=fig01.ps}
%\epsfig{file=en_sgam-200.eps,width=140mm,height=130mm}
%\end{center}
%\vspace{ -8mm}
%\baselineskip=7mm
\centering
\setlength{\unitlength}{0.1mm}
%%%%%%%%%%%%%%%%%%%%%%%%%%%%%%%%%%
%%%\begin{picture}(1600, 1540)
\begin{picture}(1600, 730)
%%\put( 450, 1530){\makebox(0,0)[cb]{\bf (a)} }
%%\put(1230, 1530){\makebox(0,0)[cb]{\bf (b)} }
%%\put(   0, 870){\makebox(0,0)[lb]{\epsfig{file=ef.ps,angle=270,
%%                                        width=80mm}}}
%%\put( 800, 870){\makebox(0,0)[lb]{\epsfig{file=hwptc.ps,angle=270,
%%                                        width=80mm}}}
%%%\put( 250, 660){\makebox(0,0)[cb]{\bf (a)} }
%%%\put(800, 660){\makebox(0,0)[cb]{\bf (b)} }
%%%\put(   0, 100){\makebox(0,0)[lb]{\epsfig{file=vgcut.eps,angle=90,
%%%                                        width=55mm}}}
%%%\put( 550, 100){\makebox(0,0)[lb]{\epsfig{file=vgcutir.eps,angle=90,
%%%                                        width=55mm}}}
\put( 450, 710){\makebox(0,0)[cb]{\bf (a)} }
\put(1230, 710){\makebox(0,0)[cb]{\bf (b)} }
\put(   0, 100){\makebox(0,0)[lb]{\epsfig{file=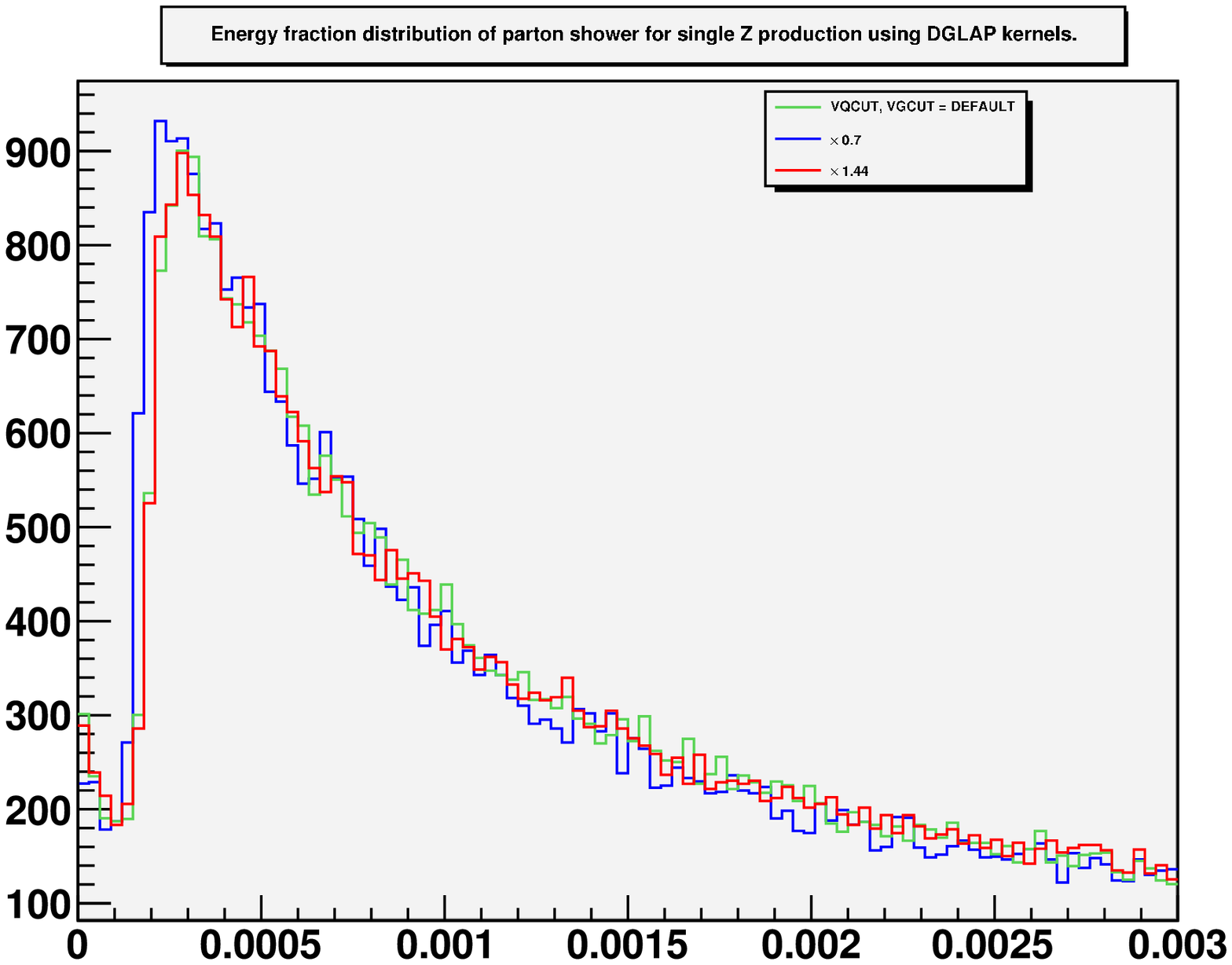,angle=90,
                                        width=75mm}}}
%\put( 800, 100){\makebox(0,0)[lb]{\epsfig{file=endglapir.ps,angle=90,
%                                        width=80mm}}}
\put( 800, 100){\makebox(0,0)[lb]{\epsfig{file=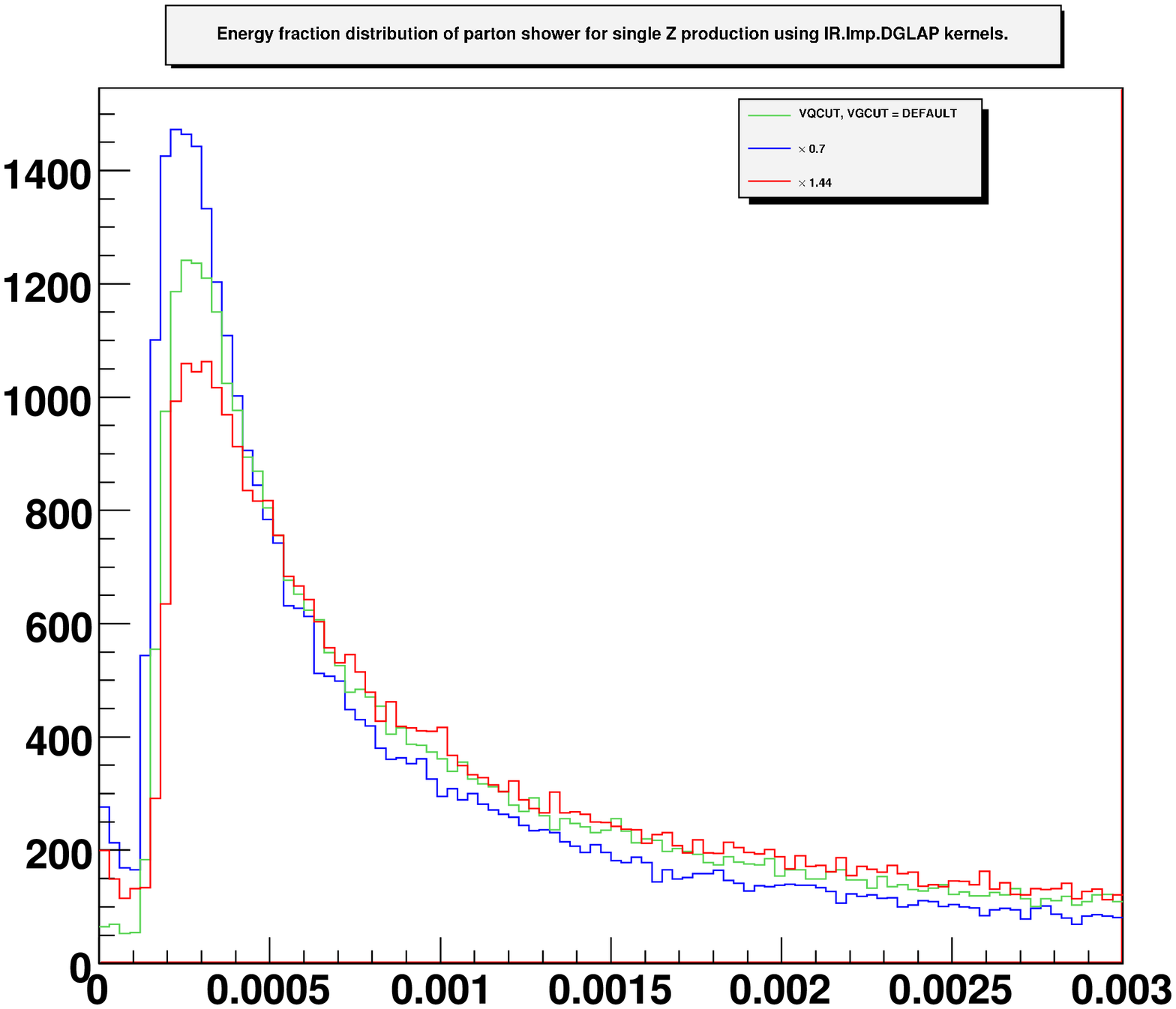,angle=270,
                                        width=70mm}}}
\end{picture}
%\vspace{ -1.5mm}
\caption{IR cut-off sensitivity in $z$-distributions of the ISR parton energy fraction: (a), DGLAP-CS 
(b), IR-I DGLAP-CS -- for the single $Z$ hard
subprocess in HERWIG-6.5 environment.
%\centerline{\Color{PineGreen}COMPARISON WITH DATA IMMINENT.} 
}
\label{fighw8}
\end{figure}

%%\begin{figure}
%%\begin{center}
%%\epsfig{file=raploose.ps,height=90mm,angle=90}
%%\end{center}
%%\caption{The $Z$ rapidity-distribution(ISR parton shower) comparison in HERWIG6.5.}
%%\label{fighw7}
%%\end{figure}
Finally, we turn to the issue of the IR cut-off in HERWIG6.5. In HERWIG6.5,
there are IR cut-off parameters used to separate real and virtual effects
and necessitated by the +-function representation of the usual 
DGLAP-CS kernels. In HERWIRI, these parameters can be taken arbitrarily close to zero, as the IR-improved kernels are integrable~\cite{irdglap1,irdglap2}\footnote{We note that in the current version of HERWIRI, the formula for $\alpha_s(Q)$ is unchanged from that in HERWIG6.5 so that there is still a Landau pole therein and this would prevent our taking the attendant IR cut-off parameters arbitrarily close to zero; however, we also note that this Landau pole is spurious and a more realistic behavior for $\alpha_s(Q)$ as $Q\rightarrow 0$ from either the lattice approach~\cite{bou} or from other approaches such as those in Refs.~\cite{shirk,max} could be introduced in the regime where the usual formula for $\alpha_s(Q)$ fails and this would allow us to approach zero with the IR cut-off parameters.}. We now illustrate the difference in IR cut-off response by comparing it for HERWIG6.5 and HERWIRI:
we change the default values of the parameters in HERWIG6.5 by factors of 0.7 and 1.44 as shown in the Fig.~\ref{fighw8}. We see that the harder cut-off reduces the phase space only significantly for the IR-improved kernels and that the softer cut-off has also a small effect on the usual kernels spectra whereas as expected
the IR-improved kernels spectra move significantly toward softer values as a convergent integral
would lead one to expect\footnote{One must note here that the spectra all stop at approximately the same value $z_0\cong .00014-.0016$ which is above some of the modulated IR-cut-off parameters, as the HERWIG environment has other 
built-in cut-offs that prevent such things as $\alpha_s$ argument's becoming 
too small. What the curves in Fig.~\ref{fighw8} show then are 
the relative ``relative'' probabilities for normalized spectra above $z_0$.}. 
This should lead to a better description of the soft
radiation data at LHC. We await confrontation with experiment accordingly.

\begin{figure}
\begin{center}
\epsfig{file=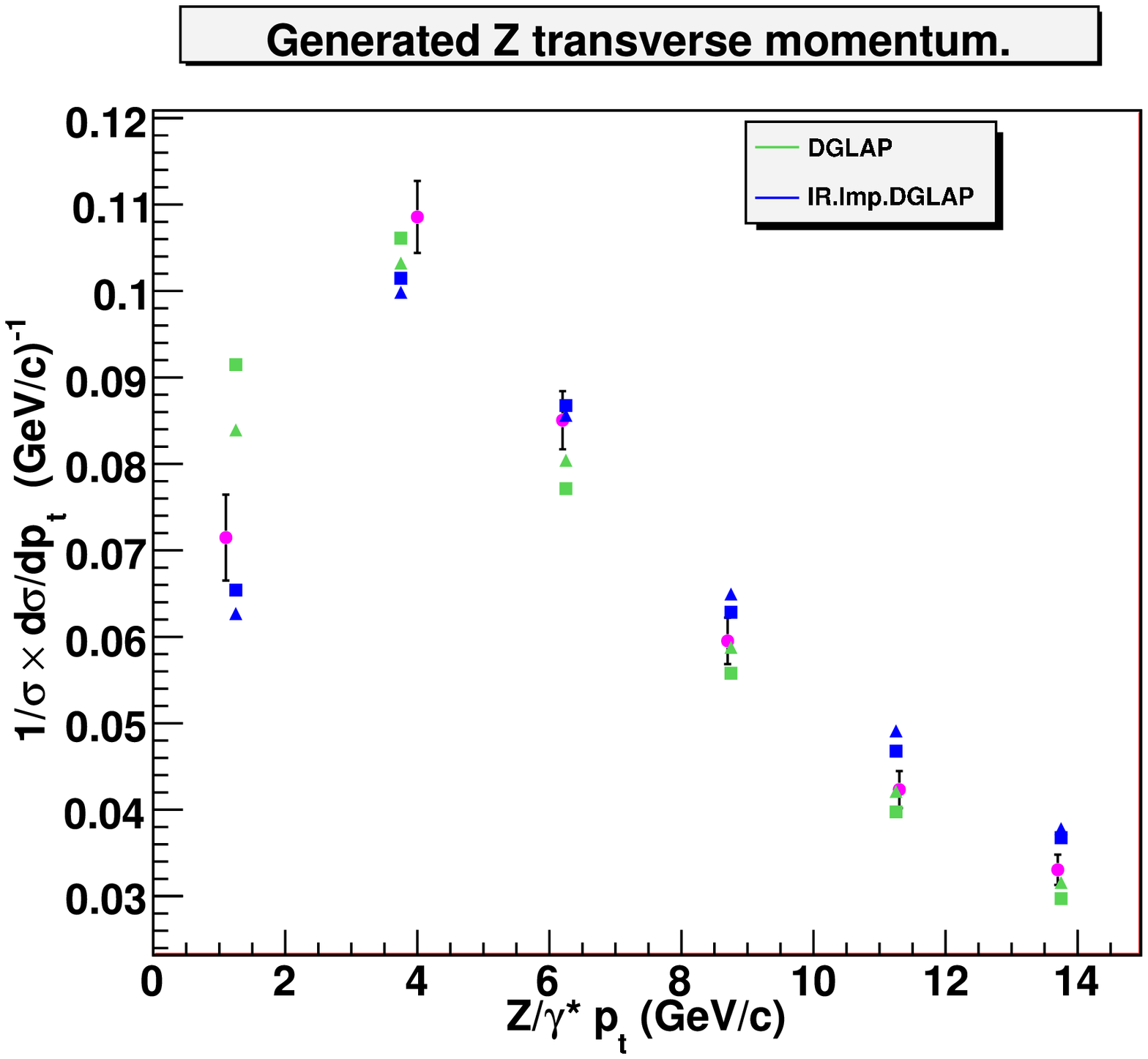,height=110mm,angle=90}
\end{center}
\caption{ Comparison with FNAL data: D0 $p_T$ spectrum data on ($Z/\gamma^*$) production to $e^+e^-$ pairs,
the circular dots are the data, the blue triangles are HERWIRI1.031, the green triangles are HERWIG6.510, the blue squares are MC@NLO/HERWIRI1.031, and the green squares are MC@NLO/HERWIG6.510.
%\centerline{\Color{PineGreen}COMPARISON WITH DATA IMMINENT.} 
}
\label{fighw9}
\end{figure}

We finish this initial comparison discussion by turning to the data from 
FNAL on the $Z$ %rapidity and 
$p_T$ spectra as reported in 
Refs.~\cite{galea,d0pt}. We show these results, for 1.96 TeV cms 
energy, in Fig.~\ref{fighw9}. 
%%We see that, in the case of the 
%%CDF rapidity data, HERWIRI1.0
%%and HERWIG6.5 both give a reasonable 
%%overall representation of the data but that
%%HERWIRI1.0 is somewhat closer to the data for small values of y. The two $\chi^2$/d.o.f are 1.76 and 1.89
%%for HERWIG6.5 and HERWIRI1.0 respectively. The data 
%%errors in Fig.~\ref{fighw9}(a)
%%do not include luminosity and PDF errors~\cite{galea}, so that
%%they can only be used conditionally at this point.
%The two $\chi^2$/p.d.f. are xxx and xxx for HERWIRI1.0 and HERWIG6.5 respectively. 
%The two
%respective $\chi^2$/d.o.f. are xxx and yyy for  HERWIG6.5 and HERWIRI1.0
%for all the data in Fig.~\ref{fighw9}(b).
%The two attendant
%$\chi^2$/p.d.f. are xxx and xxx for HERWIRI1.0 and HERWIG6.5 respectively.
%
\noindent
For these D0 $p_T$ data, we see that HERWIRI1.031 gives a better fit to the data
compared to HERWIG6.510 for low $p_T$, 
(for $p_T< 8$GeV, the $\chi^2$/d.o.f. are
$\sim$ 2.5 and 3.3 respectively if we add the statistical and systematic
errors), showing that the IR-improvement makes a better representation
of QCD in the soft 
regime for a given fixed order in perturbation theory. We have also added
the results of MC@NLO~\cite{mcnlo}\footnote{We thank S. Frixione for helpful discussion on this implementation.} for the two programs and we see
that the ${\cal O}(\alpha_s)$ correction improves the $\chi^2$/d.o.f for
the HERWIRI1.031 in both the soft and hard regimes and it improves
the HERWIG6.510 $\chi^2$/d.o.f for $p_T$ near $3.75$ GeV
where the distribution peaks.
These results are of course still subject to tuning as we indicated above.    
%For the higher values of $p_T$ both results show the need for the
%exact higher order corrections as expected, but HERWIG6.5 is
%closer to the data than is HERWIRI1.0 as we see already 

\section{Conclusions}\label{concl}
In this paper we have introduced the first QCD MC parton showers which 
do not need an IR cut-off to separate soft real and virtual corrections.
We have shown that spectra at both the parton level and at the hadron level
are softer in general. In the important process of single $Z$ production,
these IR-improved spectra show the expected behavior of an integrable
distribution. The comparison with %the CDF rapidity data and 
the D0 $p_T$ spectrum
in the soft regime shows that the IR-improvement does indeed
improve the agreement with the data. Of course, this just sets the stage for
the further implementation
of the attendant~\cite{qced} 
new approach to precision QED$\times$QCD predictions for LHC physics
by the introduction of the
respective resummed residuals needed to systematically improve
the precision tag to the 1\% regime for such processes as single heavy gauge boson production, for example. Already, however, we note that our new
IR-improved MC, HERWIRI1.031, available at http://thep03.baylor.edu, is expected to allow for a better $\chi^2$ per degree of freedom in data analysis of high energy
hadron-hadron scattering for soft radiative effects, thereby enabling a more precise comparison between theory and experiment. We have given evidence that this is indeed the case.
Accordingly, we look forward to the further exploration and development of the results
presented herein.
\section*{Acknowledgments}
%The authors wish to thank JACoW for their guidance in preparing
%this template.
%
%Work supported by Department of Energy contract DE-AC02-76SF00515.
%\end{acknowledgments}
B.F.L. Ward acknowledges helpful discussions with Prof. Bryan Webber,
Prof. M. Seymour and Prof. S. Frixione. B.F.L. Ward also thanks Prof. L. Alvarez-Gaume and Prof. W. Hollik for the support and kind hospitality of the CERN TH Division and of the Werner-Heisenberg Institut, MPI, Munich, respectively, while this work was in progress. S.A. Yost acknowledges support from The Citadel Foundation and the hospitality and support of Princeton University.

%Work partly supported by US DOE grant DE-FG02-05ER41399 and 
% the Polish Government
%grants KBN 2P30225206 and 2P03B17210, the Maria Sk\l{}odowska-Curie
%Joint Fund II PAA/DOE-97-316, and
%by NATO Grant PST.CLG.980342.

%\end{acknowledgments}

\end{document}